\documentclass[aps,prb,reprint,superscriptaddress]{revtex4-2}
\usepackage{amsfonts}
\usepackage{amsmath}
\usepackage{amssymb}
\usepackage{graphicx}
\usepackage{color}
\usepackage{amsmath}
\usepackage{physics}
\usepackage{float}
\usepackage{mathtools}
\usepackage{algorithm}
\usepackage{algpseudocode}

\usepackage[colorinlistoftodos]{todonotes}
\usepackage[colorlinks=true, allcolors=blue]{hyperref}
\usepackage{mwe}
\usepackage{epstopdf}
\usepackage[doipre={DOI:~}]{uri}

\newcommand \be {\begin{eqnarray}}
\newcommand \ee {\end{eqnarray}}
\newcommand \ben {\begin{eqnarray}}
\newcommand \een {\end{eqnarray}}

\newcommand{\kv}{\mathbf{k}}

\newcommand \rmd {\rm d}

\begin{document}
\title{Amplitude expansion of the phase-field crystal model for complex crystal structures}

\author{Marcello De Donno}
\affiliation{Institute of Scientific Computing, Technische  Universit\"at Dresden, 01062 Dresden, Germany}
\author{Lucas Benoit-{}-Maréchal}
\affiliation{Institute of Scientific Computing, Technische  Universit\"at Dresden, 01062 Dresden, Germany}
\author{Marco Salvalaglio}
\email{marco.salvalaglio@tu-dresden.de} 
\affiliation{Institute  of Scientific Computing,  Technische  Universit\"at  Dresden,  01062  Dresden,  Germany}
\affiliation{Dresden Center for Computational Materials Science (DCMS), TU Dresden, 01062 Dresden, Germany}


\begin{abstract}
The phase-field crystal (PFC) model describes crystal lattices at diffusive timescales. Its amplitude expansion (APFC) can be applied to the investigation of relatively large systems under some approximations. However, crystal symmetries accessible within the APFC model are limited to basic ones, namely triangular and square in two dimensions, and body-centered cubic and face-centered cubic in three dimensions. In this work, we propose a general, amplitudes-based description of virtually any lattice symmetry. To fully exploit the advantages of this model, featuring slowly varying quantities in bulk and localized significant variations at dislocations and interfaces, we consider formulations suitable for real-space numerical methods supporting adaptive spatial discretization. We explore approaches originally proposed for the PFC model which allow for symmetries beyond basic ones through extended parametrizations. Moreover, we tackle the modeling of non-Bravais lattices by introducing an amplitude expansion for lattices with a basis and further generalizations. We study and discuss the stability of selected, prototypical lattice symmetries. As pivotal examples, we show that the proposed approach allows for a coarse-grained description of the kagome lattice, exotic square arrangements, and the diamond lattice, as bulk crystals and, importantly, hosting dislocations. 
\end{abstract}

\maketitle

\section{Introduction}

The comprehensive modeling of crystalline materials often requires mesoscale methods, which enable the study of large systems and long timescales while retaining microscopic details. The phase-field crystal (PFC) model \cite{Elder2002,Elder2004,Emmerich2012}, or conserved Swift--Hohenberg model \cite{swift1977}, allows for studying phenomena on atomic length and relatively large (diffusive) timescales. It consists of a continuum field theory approach based on a smooth order parameter $\psi$ related to the atomic density. This model may be derived from the classical density functional theory \cite{RY79,elder2007,vanTeeffelen2009} through approximations, at the cost of accuracy loss in the description of microscopic details \cite{Oettel2012,Archer2019}. However, it has been successfully exploited to describe mesoscale phenomena such as elastic and plastic deformation in crystals, including several features and phenomena such as multicomponent systems, anisotropies, crystal growth, dislocation dynamics, and microstructure evolution \cite{Elder2004,Emmerich2012,Berry2014,Backofen14,GRANASY2019}.

The original PFC model contains two adjustable parameters only, related to the temperature and the average density. 
This poses some limitations to describing quantitative effects and complex symmetries and precludes thorough comparisons with experiments. Later formulations have been proposed to allow for such extensions. In the so-called XPFC model \cite{Greenwood2010,xtal2}, more detailed descriptions and other phenomena such as structural transformations have been achieved thanks to a definition of the free energy containing a tunable correlation function \cite{Greenwood2011binary,Berry2012,chan2015phase,alster2017simulating}, typically defined in the reciprocal (Fourier) space.
Furthermore, PFC models similar to the original formulation have been devised to account for multicomponent systems \cite{Huang2010}, complex symmetries \cite{Mkhonta2013,Backofen2021}, and quasicrystals \cite{achim2014} through extended parametrizations. Alternative approaches providing a microscopic description of complex lattice symmetries and phase transformations at diffusive timescales have also been proposed, such as the quasiparticle method \cite{lavrskyi2016quasiparticle,DEMANGE2022}.

Computing the numerical solution of the equations entering the PFC model or similar microscopic approaches inherently requires a fine spatial resolution, specifically to resolve $\psi$ on microscopic lengths, even when employing efficient numerical implementations \cite{Backofen2007,Wise2009,elsey2013simple,Yamanaka2017}. To overcome this limitation, a coarse-grained approach, namely the amplitude expansion of the PFC model (APFC), has been proposed \cite{Goldenfeld2005,Athreya2006}. In this approach, density fluctuations are described by complex amplitudes of a small set of Fourier modes reproducing the periodicity of the lattice. These amplitudes are constant in relaxed bulk crystals, vanish in the liquid phase, and can be used to characterize liquid-solid transitions. Importantly, lattice distortions and defects can be described by complex amplitudes through their phases. 
Such an amplitude formulation allows for the comprehensive characterization of the elastic field in PFC models \cite{Spatschek2010,HeinonenPRE2014,Huter2016,SalvalaglioNPJ2019}. 
Obtained as a coarse-grained PFC model, the APFC model is an approach reminiscent of classical phase-field methods that simultaneously includes interfaces, lattice deformations, defects, and retains details of the underlying lattice structure. As such, it is suitable for microscopically informed mesoscale investigations approaching continuum lengthscales  \cite{SalvalaglioPRL2021,Jreidini2021}. Amplitudes oscillate significantly at defects and interfaces, while elsewhere they vary slowly compared to lattice periodicity. Therefore, numerical approaches featuring inhomogeneous spatial discretization have been developed \cite{AthreyaPRE2007,Bercic2018,Praetorius2019} and allowed the study of mesoscopic lengthscales, also for three-dimensional systems \cite{SalvalaglioNPJ2019}. For a recent review, see Ref.~\cite{SalvalaglioMSMSE2022}.

To date, the APFC model has been exploited for simple lattice symmetries only, namely triangular and square lattices in two dimensions as well as body-centered cubic and face-centered cubic (fcc) in three dimensions. In an attempt to model
more complex symmetries, an amplitude formulation of the XPFC model has been proposed \cite{Ofori-Opoku2013} that exploits the same concept of tunable correlation functions as the original XPFC model. However, while in principle this formulation allows for modeling complex structures, it has so far only been applied to the study of simple symmetries in two dimensions. More importantly, numerical simulations based on this model rely on spectral (Fourier) methods. Although feasible, simulations are then generally limited to relatively small scales, in particular for three-dimensional investigations. Indeed, efficient numerical approaches exploiting spectral methods must feature a uniform spatial resolution that resolves variables at defects and interfaces, thus approaching the one required by the PFC model in the first place. 

In this work, we provide a general framework for modeling lattice symmetries within the APFC model, suitable for real-space numerical methods. We begin with the basics of the PFC and APFC model in Sec.~\ref{sec:models}. In Sec.~\ref{sec:newmodels} we then introduce novel model extensions together with their assessment through proofs of concept and comparisons to known results. In particular, we present the multimode amplitude expansion for the PFC model (Sec.~\ref{sec:3a}), the amplitude expansion featuring a local structure on Bravais lattice sites (Sec.~\ref{sec:3b}), and an additional stabilizing term filtering out nonphysical phases in solid-state crystalline systems (Sec.~\ref{sec:3c}). In Sec.~\ref{sec::numerical} we then apply the framework, providing novel mesoscale simulations of crystal structures hosting dislocations, and showing the capabilities of the model in both two and three spatial dimensions. A few representative cases are addressed, such as the kagome lattice, different triangular and square lattices with a basis, and the diamond lattice.
We finally draw our conclusions and remarks in Sec.~\ref{sec:conclusions}.

\section{PFC and APFC Model}
\label{sec:models}
The PFC model is based on the Swift--Hohenberg energy functional, which can be written as \cite{Elder2002,Elder2004,Emmerich2012}
\begin{equation}
\begin{split}
F_\psi=\int_{\Omega} \left[\frac{A}{2} \psi\mathcal{L}\psi + \frac{B}{2}\psi^2
+\frac{C}{3}\psi^3+\frac{D}{4}\psi^4 \right] \rmd\mathbf{r},
\label{eq:F_PFC}
\end{split}
\end{equation}
where $\mathcal{L}=(q^2+\nabla^2)^2$, $A$, $B$, $C$, and $D$ are parameters \cite{elder2007} and $q$ is the length of the principal (shortest) wave vector for periodic minimizers of $F_\psi$. Under the assumption of dissipative dynamics driven by the minimization of $F_\psi$ conserving the order parameter, the evolution law for $\psi$ is given by
\begin{equation}
\frac{\partial \psi}{\partial t}=\nabla^2\frac{\delta F_\psi}{\delta \psi}.
\label{eq:dpsidt}
\end{equation}
This approach allows for describing crystalline systems over diffusive timescales \cite{Emmerich2012}. Different dynamics and extensions have been proposed to provide advanced modeling of elastic relaxation \cite{StefanovicPRL2006,tothNonlinearHydrodynamicTheory2013,SkaugenPRL2018,SkogvollHPFC2022}.

The coarse graining achieved by the APFC model is based on the approximation of $\psi$ by a sum of plane waves with complex amplitudes $\eta_j$ \cite{Goldenfeld2005,Athreya2006,SalvalaglioMSMSE2022}: 
\begin{equation}
    \psi(\mathbf{r}) = \bar{\psi} + \sum_{n=1}^N \eta_n e^{{\rm i}\kv_n\cdot \mathbf{r}}+\text{c.c.}=
    \sum_{n=-N}^N  \eta_n e^{{\rm i}\kv_n\cdot \mathbf{r}},
    \label{eq:n_app}
\end{equation}
where $\eta_{-n}=\eta^*_n$, $\eta_0=\bar{\psi}$ is assumed to be constant for the purposes of this work, $\kv_n$ are the reciprocal space vectors reproducing a specific lattice symmetry \cite{SalvalaglioMSMSE2022}, with $\kv_{-n}=-\kv_n$, $\kv_0=\mathbf{0}$, and c.c. referring to the complex conjugate. Usually a small set of modes, with ``mode" referring to a family of equal-length $\kv_n$, are considered. 

The APFC model describes crystalline phases by focusing on slowly varying complex amplitudes. A free-energy functional $F_\eta$ depending on $\eta_n$ can be formally obtained through a renormalization group approach \cite{Goldenfeld2005} or, equivalently, by substituting Eq.~\eqref{eq:n_app} into \eqref{eq:F_PFC} and integrating over the unit cell under the assumption of constant amplitudes therein \cite{Athreya2006,SalvalaglioMSMSE2022}. For a one-mode approximation of $\psi$ with $|\kv_n|=q$, the free energy reads
\begin{equation}
F_\eta
=\int_{\Omega} \left[ \sum_{n=1}^N \left( A^\prime|\mathcal{G}_{n}\eta_n|^2 \right) +g^{\rm s}(\{\eta_n\})  \right]\ \rmd\mathbf{r}, 
\label{eq:APFC}
\end{equation}
with $\mathcal{G}_{n}=\nabla^2+2{\rm i}\kv_n\cdot\nabla$, and $g^{\rm s}$ 
a polynomial in the amplitudes whose terms depend on the lattice symmetry. It takes the form:
\begin{equation}\label{eq:gpar}
    g^{\rm s}
    =\frac{B^\prime}{2}\zeta_2
    +\frac{C^\prime}{3}\zeta_3
    +\frac{D^\prime}{4}\zeta_4
    +E^\prime, 
\end{equation}
where
\begin{equation}
\begin{split}
    \zeta_2=&\sum_{p,q\in I(N)}\eta_p\eta_q \delta_{\mathbf{0},\kv_p+\kv_q}
    =2\sum_{n=1}^N |\eta_n|^2=\Phi,
    \\
    \zeta_3=&\sum_{p,q,r\in {I}(N)}\eta_p\eta_q\eta_r \delta_{\mathbf{0},\kv_p+\kv_q+{\kv_r}},
    \\
    \zeta_4=&\sum_{p,q,r,s\in {I}(N)}\eta_p\eta_q\eta_r\eta_s \delta_{\mathbf{0},\kv_p+\kv_q+{\kv_r}+\kv_s},
    \label{eq:pol_resonance}
\end{split}
\end{equation}
with $I(N):=\{ i\in \mathbb{Z}:0<|i|\leq N \}$, and $\delta_{\mathbf{0},\mathbf{Q}}=1$ if $\mathbf{Q}=\mathbf{0}$ and 0 elsewhere. The coefficients relate to those entering Eq.~\eqref{eq:F_PFC} as follows: $A^\prime=A$, $B^\prime=B+2C\bar{\psi}+3D\bar{\psi}^2$, $C^\prime=C+3D\bar{\psi}$, $D^\prime=D$, and $E^\prime=(A/2)\bar{\psi}^2+(C/3)\bar{\psi}^3+(D/4)\bar{\psi}^4$. Explicit forms of $g^{\rm s}$ can be found in Ref.~\cite{SalvalaglioMSMSE2022}. 
The dynamics of the amplitudes approximating Eq.~\eqref{eq:dpsidt} is given by \cite{Goldenfeld2005,Athreya2006}
\begin{equation}
\frac{\partial \eta_n}{\partial t}=-|\kv_n|^2\frac{\delta F_{\eta}}{\delta \eta_n^*}=-|\kv_n|^2\left(A^\prime\mathcal{G}_n^2\eta_n+\frac{\partial g^{\rm s}}{\partial \eta_n^*}\right),
\label{eq:detadt}
\end{equation}
which may be supplied by a conservative dynamic for the average density \cite{Yeon2010} and further extensions to account for proper elastic relaxation \cite{HeinonenPRE2014,HeinonenPRL2016,SalvalaglioJMPS2020}.

Simulations reported in the following are performed exploiting the adaptive Finite Element Method toolbox AMDiS \cite{Vey2007,WitkowskiACM2015}. Such an approach, described in detail in Refs.~\cite{Salvalaglio2017,SalvalaglioNPJ2019,Praetorius2019}, allows us to efficiently solve the APFC equations with optimized mesh refinement. Minor changes have been implemented to include the extensions reported in the next section.

\begin{figure*} 
    \includegraphics[width=\linewidth]{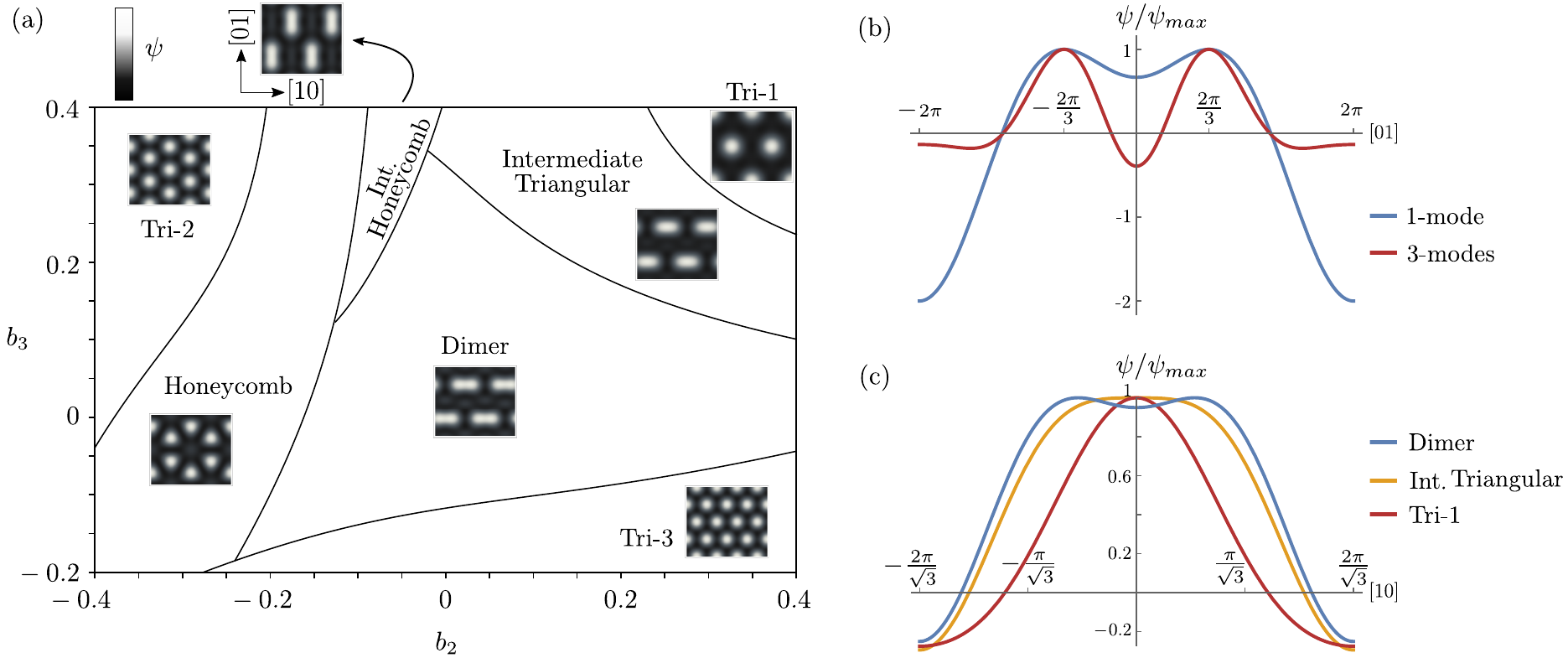}
    \caption{Multimode APFC model encoding a three-mode approximation of a triangular lattice (see definition of $\mathbf{k}_n$ in Appendix \ref{app:triangular}). (a) Phases obtained by varying $b_{2,3}$ with $b_1=0$, $\bar{\psi} = 0.2,\,A^\prime = 0.02,\,B' = - 0.15,\,C' = 0,\,D' = 1$. Density reconstructed from Eq.~\eqref{eq:n_app} are shown as insets. 
    (b) One-dimensional density profile along the $[01]$ direction, for the honeycomb structure as from panel (a) with $b_2=-0.2$, $b_3=0.2$, compared to the same quantity obtained for a one-mode approximation. 
    (c) One-dimensional density profile along the $[10]$ direction for the dimer ($b_2=0$, $b_3=0.$), intermediate triangular ($b_2=0.2$, $b_3=0.3$), and triangular  (Tri-1, $b_2=0.4$, $b_3=0.4$) structures.} 
    \label{fig:graphene_bi}
\end{figure*}

\section{APFC modeling of complex symmetries}
\label{sec:newmodels}

\subsection{Multimode approximation}\label{sec:3a}

One-mode approximations, as considered in Eq.~\eqref{eq:APFC}, only allow for a few crystal symmetries. Even the basic fcc lattice symmetry is typically described by the APFC model only when considering a two-mode approximation \cite{ElderPRE2010,Salvalaglio2017}. To extend the APFC modeling towards complex lattice symmetries, we consider the amplitude formulation for an arbitrary number of modes by exploiting the multimode Swift--Hohenberg free-energy functional \cite{Mkhonta2013}.
In this approach, the differential operator $\mathcal{L}$ entering the free energy \eqref{eq:F_PFC} may be written as
\begin{equation}
    \mathcal{L}_{M}= \prod_{m=1}^{M} \mathcal{L}_m= \prod_{m=1}^{M}\left[(q_m^2+\nabla^2)^2+b_m\right], 
    \label{eq:LM}
\end{equation}
where the parameters $b_m$ control the relative stability of the corresponding modes at wave numbers $q_m$. For $M=1$, $q_1\equiv q$ and $b_1=0$, we recover $\mathcal{L}_{M}=\mathcal{L}_{1}=\mathcal{L}$ as appearing in Eq.~\eqref{eq:F_PFC}. Such an energy term is the generalization of the gradient term in two-mode PFC or Swift--Hohenberg models, that has been successfully exploited to model diverse crystal symmetries such as fcc, square and quasiperiodic lattices \cite{Lifshitz1997,Wu2010}. Although not rigorously derived from microscopic descriptions, it phenomenologically encodes the stability of the modes with lengths $q_m$, which is the desired property to have an energy functional minimized by Eq.~\eqref{eq:n_app} with prescribed $\mathbf{k}_n$ as set in the APFC model.
The corresponding operator to be considered in Eqs.~\eqref{eq:APFC}--\eqref{eq:detadt} can be derived in a similar fashion to the other terms in the amplitude equations  \cite{Goldenfeld2005,Athreya2006,SalvalaglioMSMSE2022}. From Eq.~\eqref{eq:LM} we obtain
\begin{equation}
    \begin{split}
        \mathcal{M}_{\mathrm{M},n}=& \prod_{m=1}^{M} \left[(\mathcal{G}_n-|\kv_n|^2+q_m^2)^2+b_m\right]\\
        =&\, (\mathcal{G}_n^2 + b_n) \prod_{m\neq n}^{M} \left[(q_m^2 - |\kv_n|^2+\mathcal{G}_n)^2+b_m\right])\\
        \approx&\,  (\mathcal{G}_n^2 +b_n)
        \underbrace{\prod_{m\neq n}^{M} \left[(q_m^2 - |\kv_n|^2)^2 + b_m \right]}_{\Gamma_n}\\
        \equiv&\, \Gamma_n \left[(\nabla^2+2{\rm i}\kv_n\cdot\nabla)^2+b_n\right],
        \label{eq:MM}
    \end{split}
\end{equation}
where the considered approximation holds true in the limit $2|\mathcal{G}_n\eta_n|\ll|(q_m^2 - |\kv_n|^2)\eta_n|$, valid for slowly oscillating amplitudes analogously to the basic assumptions underlying the APFC model \cite{SalvalaglioMSMSE2022}. In this case, for $M=1$, $q_1\equiv q$, and $b_1=0$ we have that $\mathcal{M}_{{1},n}=\mathcal{G}^2_n$ enters \eqref{eq:detadt}.
The operator \eqref{eq:MM} leads to an additional contribution to $E^\prime$ entering Eq.~\eqref{eq:gpar}, now reading
\begin{equation}\label{density_E}
    E^\prime = \frac{A}{2}\prod_{m=1}^M(|\mathbf{k}_m|^4+b_m)\bar{\psi}^2+
    \frac{B}{2}\bar{\psi}^2+
    \frac{C}{3}\bar{\psi}^3 + 
    \frac{D}{4}\bar{\psi}^4.
\end{equation}

As a prototypical example, we focus on the three-mode approximation of a triangular lattice. 
The specific choice of $\mathbf{k}_n$ is reported in Appendix \ref{app:triangular}. At this stage, we are interested in relaxed crystal phases. Therefore we can numerically minimize the free energy $F_\eta$ with respect to constant and real amplitudes $\eta_n\equiv \phi_n$. Figure~\ref{fig:graphene_bi}(a) shows a phase diagram obtained by minimizing the energy with $b_1 = 0$ and varying $b_2$ and $b_3$.
Corresponding reconstructed densities from Eq.~\eqref{eq:n_app} are reported in insets. While retaining the underlying triangular symmetry, different structures are obtained from different superpositions of Fourier modes in Eq.~\eqref{eq:n_app} weighted by the (real) amplitudes ensuring energy minimization. They consist of triangular phases with different lattice spacing as dictated by the different sets of reciprocal-space vectors, as well as a honeycomb and a dimerlike phase. Triangular and honeycomb lattice symmetries can be described with a one-mode approximation \cite{Athreya2006,ElderPRE2010,Salvalaglio2017}. However, the extended set of $\mathbf{k}_n$ allows for a better resolution of the density peaks, as shown in Fig.~\ref{fig:graphene_bi}(b).  Therefore, it can be exploited to account for additional small-scale details while solving equations for slowly varying oscillating fields. Intermediate triangular and honeycomb structures are also obtained. The former corresponds to a triangular arrangement of asymmetric peaks (see also Fig.~\ref{fig:graphene_bi}(c)), the latter to the arrangement of six maxima qualitatively similar to a honeycomb structure but with varying spacing. 

Interestingly, the results reported in Fig.~\ref{fig:graphene_bi} show a good agreement with an analogous investigation performed with the multimode PFC model in Ref.~\cite{Mkhonta2013}, i.e., with Eqs.~\eqref{eq:F_PFC} and \eqref{eq:dpsidt} and $\mathcal{L}=\mathcal{L}_M$ as from Eq.~\eqref{eq:LM}. Therein, three lengthscales $q_{1,2,3} = 1,\sqrt{3}, 2$ were considered. This choice enforces the length of the shortest wave vector, similarly to the APFC model considered here. In the PFC model,  however, there is no restriction of the solution to a prescribed set of $\mathbf{k}_n$ vectors. In other words, higher harmonics may be present as well as further nontrivial combinations of modes, meaning that the solution may be described by subsets of all possible reciprocal-space vectors of given lengths. As a result, a slightly larger set of lattices is observed. However, phases with an underlying triangular symmetry are obtained for very similar values of $b_{2,3}$.

\begin{figure}
    \centering
    \includegraphics[width=\linewidth]{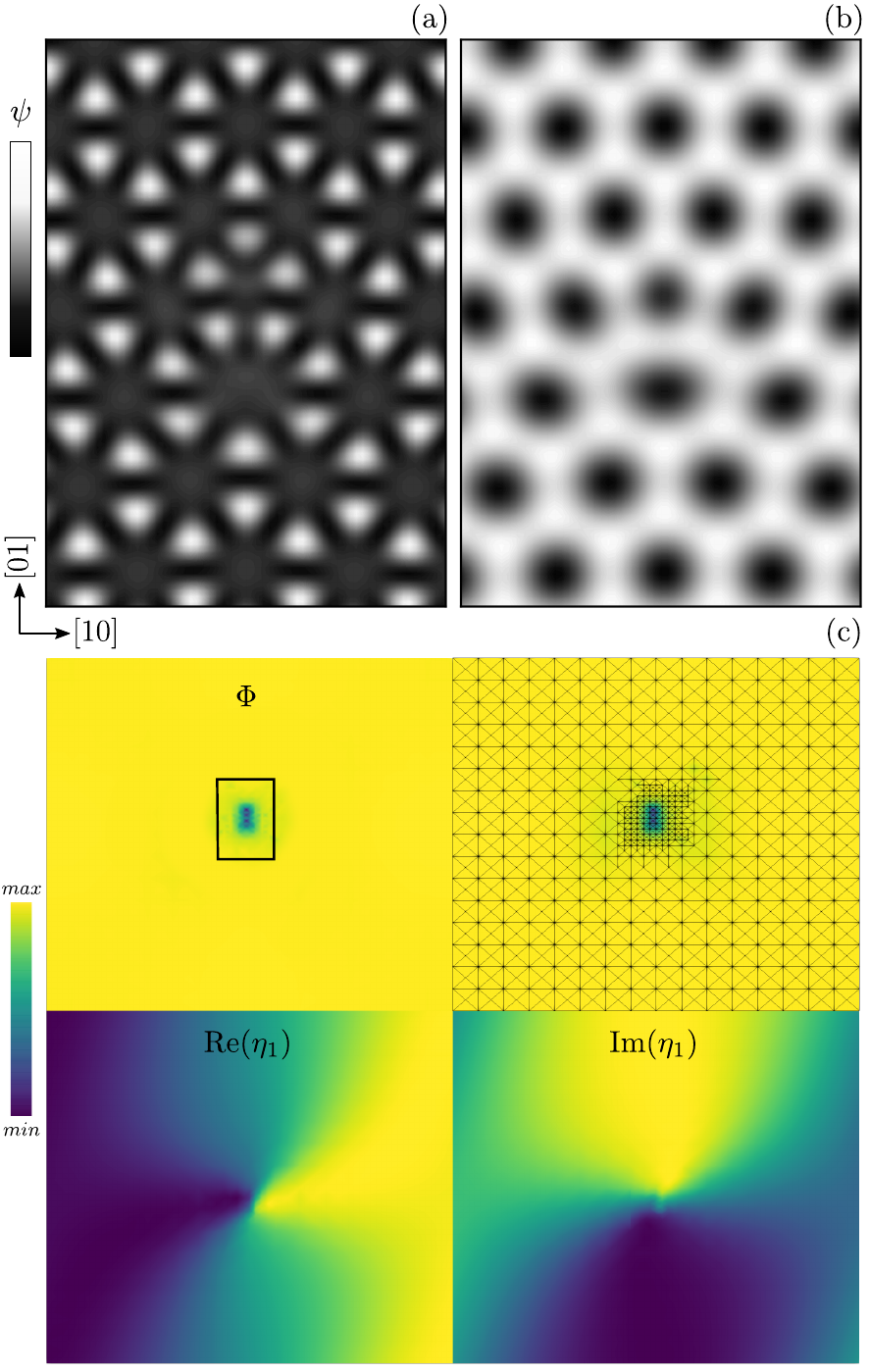}
    \caption{
    Edge dislocation in a honeycomb lattice. Plots of the reconstructed density $\psi(\mathbf{r})$ are shown for the following cases: 
    (a) A three-mode approximation, $\bar{\psi} = 0,\,A=0.02,\, B=0.02,\, C=-0.5,\, D=1/3,\, b_1=0.1,\, b_2=b_3=0$. 
    (b) A one-mode approximation, $\bar{\psi} = 0,\,A=1,\, B=0.02,\, C=-0.5,\, D=1/3,\,b_{n}=0$. 
    (c) Representation of the entire simulation domain for the three-mode approximation. Relevant quantities are plotted in different quadrants. Top left: $\Phi = 2\sum_{n=1}^N |\eta_n|^2$; the black rectangle shows the size of the region reported in panel (a). Top right: $\Phi$ superimposed to the locally refined simulation grid. Bottom: Real and imaginary part of $\eta_1$.}
    \label{fig:graphene_defects}
\end{figure}

The phases described in Fig.~\ref{fig:graphene_bi} can be considered by the multimode APFC model also when dealing with deformed crystals hosting defects.
Figure~\ref{fig:graphene_defects}(a) illustrates the 5$|$7 defect core of a point edge dislocation in a honeycomb lattice. The reconstructed density from Eq.~\eqref{eq:n_app} with amplitudes computed by three-mode APFC approximation as discussed above is shown. The dislocation is obtained at the interface between layers with different numbers of lattice planes following a conventional approach from the literature \cite{Salvalaglio2017}, mimicking a system of layers with opposite strain along the interface. This is encoded in the initial condition by
\begin{equation}\label{eq:inieta}
    \eta_n = \phi_n e^{-\mathrm{i} \mathbf{k}_n\cdot\mathbf{u}},
\end{equation}
with $\phi_n$ the value of the amplitude in the relaxed bulk as above, and $\mathbf{u}=(\pm \epsilon x,0)$ with strain $\epsilon = 1/64$. For comparison, the same defect in the honeycomb lattice obtained by a one-mode approximation is reported in Fig.~\ref{fig:graphene_defects}(b), showing the same defect structure with a coarser resolution. 

It is worth recalling that the variables to solve for in the APFC model are the complex amplitudes, varying on lengthscales larger than the reconstructed atomic density, while fully encoding lattice deformation of isolated defects. Figure~\ref{fig:graphene_defects}(c) shows the entire simulation domain, containing four defects in a static configuration, from which Fig.~\ref{fig:graphene_defects}(a) was extracted. The domain is divided into four quadrants, containing a dislocation each, in which we illustrate $\Phi = 2\sum_{n=1}^N |\eta_n|^2$, the optimized mesh exploited for simulations (see Ref.~\cite{Praetorius2019} for details on the specific criteria), as well as the real and imaginary part of $\eta_1$.
$\Phi$ plays the role of an order parameter for the crystalline phase, being constant in the bulk and decreasing at the defects. Importantly, as the amplitudes $\eta_n$ vary slowly compared to the atomistic lengthscales---with the largest gradients occurring at the dislocations \cite{Athreya2006,Goldenfeld2005}---real-space methods with inhomogeneous spatial discretization can be advantageously used to solve the equations of the APFC model, compared to reciprocal space methods \cite{SalvalaglioMSMSE2022}.
Additionally, $\eta_n$ may be exploited to extract the strain and stress fields, preserving some details of the microscopic scales together with a convenient regularization of these fields at the core  \cite{SalvalaglioNPJ2019,SalvalaglioJMPS2020}. While the goal of using the APFC model remains to describe large lengthscales with continuous fields, the reconstructed core matches almost perfectly the one obtained by the three-mode PFC model reported in Ref.~\cite{Hirvonen16}. Accordingly, information on small scales may be used to assess the present descriptions, as will also be exploited in the following.
It is worth mentioning that the operator $\mathcal{M}_{M,n}$ contributes to the solution shown in Fig.~\ref{fig:graphene_defects}(a) through the parameters $b_n$---like in the bulk system case discussed in Fig.~\ref{fig:graphene_bi}---and through the differential operators---owing to varying complex amplitudes in deformed lattices hosting dislocations. Therefore, the good agreement with the corresponding multimode PFC counterpart also represents a further assessment of the approximation considered in Eq.~\eqref{eq:MM}. Importantly, this approximation leads to a convenient form for numerical approaches, featuring similar terms to the classical, one-mode APFC model, with additional amplitude-dependent factors only and without increasing the order of the differential equations to solve.

\subsection{Amplitude expansion for lattices with a local structure}\label{sec:3b}

APFC models rely on the definition of $\kv_n$ vectors reproducing Bravais lattices. 
To account for a broader set of lattice symmetries, we introduce an amplitude expansion accounting for a local structure on Bravais-lattice sites, 
\begin{equation}\label{eq:n_app_basis}
    \psi(\mathbf{r}) -\bar{\psi} = \sum_{n=1}^N  \underbrace{\eta_n \mathcal{B}_n}_{\widetilde{\eta}_n} e^{{\rm i}\kv_n\cdot \mathbf{r}} +\text{c.c.},
\end{equation}
with $\mathcal{B}_n$ a sum of terms encoding phase shifts for the Bravais lattice periodicity and $\widetilde{\eta}_n= \eta_n \mathcal{B}_n$ modified amplitude functions. Specifically, we describe a lattice with a basis including $J$ atoms by

\begin{equation}\label{eq:basis}
    \mathcal{B}_n = \sum_{j=1}^{J} e^{-\mathrm{i} \mathbf{k}_n \cdot \mathbf{R}_j},
\end{equation}
where $\mathbf{R}_j$ is the position of the $j$th atom within the unit cell. In the case of a Bravais lattice, $J=1$ and we may set $\mathbf{R}_1=\mathbf{0}$ without loss of generality, so that $\mathcal{B}_n = 1$ and Eq.~\eqref{eq:n_app_basis} reduces to Eq.~\eqref{eq:n_app}. In the case of a lattice with a basis (same group of atoms per Bravais-lattice site), $\mathbf{R}_j$ can be set to the distance from a reference position within the unit cell.

By considering the derivation of the amplitude equations with the ansatz \eqref{eq:n_app_basis}, and also accounting for the multimode formulation in \eqref{eq:MM}, we obtain the following free energy

\begin{equation}\label{eq:F_full}
    \begin{split}
        F_{\widetilde{\eta}}=\int_{\Omega}
            &\left[\frac{A}{2} \sum_{n=1}^N \right.
            \left(\widetilde{\eta}_n \mathcal{M}_{{\rm M},n}\widetilde{\eta}_n^*
            +\widetilde{\eta}_n^* \mathcal{M}_{{\rm M},n}\widetilde{\eta}_n\right)+\\
            &\left.\phantom{\sum_1^N}+\frac{B}{2}\widetilde{\Phi}+\frac{C}{3}\zeta_3(\{\widetilde{\eta}_n\}) +\frac{D}{4}\zeta_4(\{\widetilde{\eta}_n\})
            \right] \rmd\mathbf{r},
    \end{split}
\end{equation}
with $\widetilde{\Phi} = 2\sum_{n=1}^N |\widetilde{\eta}_n|^2$, that implicitly depends on $\mathcal{B}_n$ via the definition of $\widetilde{\eta}_n$ from Eq.~\eqref{eq:n_app_basis}. Notice that the coefficients $\mathcal{B}_n$ appear as factors of terms present in Eq.~\eqref{eq:APFC}, and the resonance conditions in \eqref{eq:pol_resonance} readily apply to the products of different $\mathcal{B}_n$ factors. Like the free energy considered in the classical APFC model, $F_\mathbf{\tilde{\eta}}$ is rotationally invariant. A Bravais lattice symmetry can then be described through Eq.~\eqref{eq:n_app_basis} by a proper set of $\mathbf{k}_n$, while a local structure is encoded by $\mathcal{B}_n$. A central point that remains to address is whether the lattices described by such an approach correspond to the global minimum of $F_\mathbf{\tilde{\eta}}$ for a given set of parameters. 

The honeycomb lattice is a lattice with a basis. It represents a good test for the newly introduced formulation as it is known to be a global minimum of the (A)PFC free energies for some parameters. In the one-mode approximation, it can be described by Eq.~\eqref{eq:n_app_basis} with $N=3$, $\mathbf{k}_n$ as in Eq.~\eqref{eq:kvec_tri_1mode}, $J=2$, $R_1=(0,0)$, $R_2=(0,4\pi/3)$, and $\mathcal{B}_n=\frac{1}{2}+i\frac{\sqrt{3}}{2}$. The free-energy density for the corresponding bulk system reads
\begin{equation}
    \begin{split}
        \frac{F_{\widetilde{\eta}}}{V} =& B(\phi_1^2+\phi_2^2+\phi_3^2) + 2sC\phi_1\phi_2\phi_3+\\
            & + 3D(\phi_1^2+\phi_2^2+\phi_3^2)^2 - \frac{3}{2}D(\phi_1^4+\phi_2^4+\phi_3^4),
    \end{split}
\end{equation}
where $\phi_n$ is the bulk value of the $n$th amplitude. Notably, quadratic and quartic terms in $\mathcal{B}_n$ reduce to 1, and the entire contribution of the basis coefficients is then contained in a real constant $s$ multiplying the cubic term. It reads:

\begin{equation}
    s=\mathcal{B}_1\mathcal{B}_2\mathcal{B}_3+\rm{c.c.}=
        \begin{cases}
            -2 \quad\mathrm{honeycomb},\\
            +2 \quad\mathrm{triangular},
        \end{cases}
\end{equation}
with the triangular lattice corresponding to $\mathcal{B}_n=1$, i.e., without a basis, as described above. 
Including the basis in a one-mode approximation of the triangular lattice in order to obtain the honeycomb structure results then in an opposite sign for the cubic term. This is in full agreement with a well-known result of (A)PFC models for the one-mode triangular lattice: Free energies with opposite signs of the $C$ coefficient are minimized by $\psi_+$ (triangular phase) and $\psi_-$ (honeycomb phase) with $\psi_-=-\psi_+$ and $F_\eta(\psi_-)=F_\eta(\psi_+)$. 
The honeycomb lattice can then be obtained from a one-mode approximation of a triangular Bravais lattice either by properly setting $C$ for an amplitude expansion based on Eq.~\eqref{eq:n_app} or by considering a basis through Eq.~\eqref{eq:n_app_basis}. The latter way is preferable, as it encodes information on the target phase directly in a physically meaningful way: setting a basis on a Bravais lattice. As will be discussed in the following, this is crucial when considering lattice symmetries which can be obtained neither from a combination of modes for Bravais lattices encoded in Eq.~\eqref{eq:n_app}, nor by tuning the parameters.

This argument can be further appreciated by looking at a similar analysis performed on the three-mode approximation of a bulk honeycomb crystal. In an appropriate parameter range and without including a basis, real amplitudes within the same mode are equal in modulus, but not in their signs. That is, the energy is minimized by $\phi_n$ with
\begin{equation}
    \begin{split}
        \phi_1=-\phi_2=\phi_3&>0,\\
        -\phi_4=\phi_5=-\phi_6&>0,\\
        \phi_7=\phi_8=\phi_9&<0.
    \end{split}
\end{equation}
The honeycomb structure is then obtained with nontrivial combinations of Fourier modes in Eq.~\eqref{eq:n_app}. Indeed, the lattice symmetry imposed by $\mathbf{k}_n$ is the triangular one, while the honeycomb happens to be compatible with a different superposition of the considered Fourier modes. This loss of correspondence between the structure imposed in the amplitude expansion and the final density may prevent the tuning of lattice features, such as the resolution of the peaks, and the exploration of symmetry breaking and lattice anisotropy when considering additional physical contributions (see for example the coupling with magnetic field in Ref.~\cite{Backofen2022}). 
When introducing the basis, the honeycomb structure is obtained when all amplitudes are positive, and those within the same mode are equal, meaning that the target structure is obtained with the simplest possible combination of Fourier modes. 
Similarly to the one-mode approximation, the density profile is identical with and without the basis, as is the corresponding energy. Indeed, the amplitude expansion including the basis may be considered a different representation of densities that results from the same energy functional \eqref{eq:F_PFC}. Still, this representation allows exploring a larger set of possible symmetries. For instance, it will be shown in Sec. \ref{sec::numerical} that considering a basis is necessary to describe some crystalline structures.  

\subsection{Stabilization}\label{sec:3c}

\begin{figure}
    \centering
    \includegraphics[width=\linewidth]{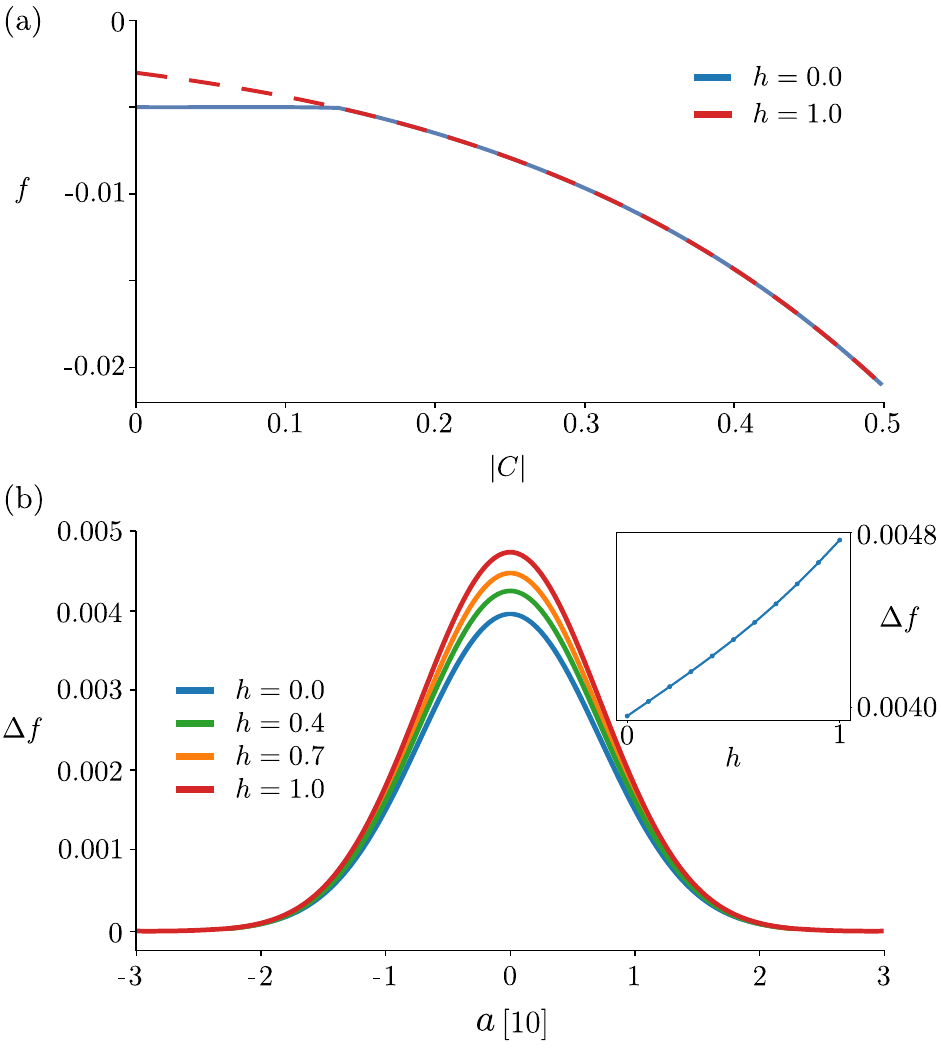}
    \caption{Effect of the stabilizing term in a one-mode triangular lattice. (a) Minimum of the free-energy density plotted against the parameter $C$, with ($h=1$, red, dashed curve) and without (blue, solid curve) the stabilizing term. (b) Maximum of the free-energy difference at a dislocation in a systems as in Fig.~\ref{fig:graphene_defects} for different values of $h$, plotted along the crystallographic direction [10], expressed in units of the lattice parameter $a=4\pi/\sqrt{3}$. Inset: Free-energy difference between defect and bulk for increasing values of the coefficient $h$.}
    \label{fig:stabterm}
\end{figure}

As discussed in the previous section, we are interested in lattices for which, in bulk and in the absence of further symmetry breaking, Fourier modes with the same $|\mathbf{k}_n|$ have the same coefficient $\eta_n$.   
However, these phases may be in competition with other phases for which subsets of the amplitudes vanish. A prominent example is the so-called stripe phase \cite{Elder2002,Elder2004}, 
which might be used to model smectics \cite{Elder1992, ElderPRA1992, Praetorius2018, HuangCP2022} but can be considered nonphysical for solid-state crystals. Some DFT formulations based on $\ln(\psi)$ rather than $\psi$ have been shown to avoid lamellar/stripe phases as global minimum of the corresponding energy \cite{Subramanian2021}.
Here, in order to filter out these cases in the APFC model, and therefore stabilize targeted phases set by an ansatz as in Eq.~\eqref{eq:n_app_basis}, we instead consider an additional energy term
$\int_\Omega f_{\rm stab}{\rm d}\mathbf{r}$ with
\begin{equation}\label{eq:stabterm}
    f_{\rm stab}=\sum_{m=1}^M \left[ h_m \sum_{j,i>j}^{N} \left(|\mathcal{B}_i|^2|\eta_i|^2-|\mathcal{B}_j|^2|\eta_j|^2\right)^2\right].
\end{equation}
This purely phenomenological term penalizes differences in the amplitudes of the same family. Note, however, that it does not alter the bulk energy of phases where these amplitudes have the same value. A demonstration is given in Fig.~\ref{fig:stabterm}(a) for a one-mode approximation of a triangular lattice ($M=1$, $h_m=h_1=h$). We show the minimum of the free-energy density by varying $C$, with ($h=1$) and without ($h=0$) the stabilizing term. 
In the latter case, the energy is minimized by a stripe phase for $|C| \lesssim 0.13$, where only one amplitude is different from zero and therefore the cubic term vanishes, rendering the energy independent on $C$. For $|C| \gtrsim 0.13$, the energy is minimized by a triangular phase. 
With $h=1$ the energy is always minimized by a triangular phase as the newly introduced term penalizes the stripe phase. Notice that the two curves overlap when the triangular phase is stable in both settings, showing explicitly that the energy of the relaxed bulk triangular phase is not affected by the additional energy term.

Amplitudes vary differently at defects \cite{SalvalaglioMSMSE2022}. To evaluate the effect of the stabilizing term, we consider a system analogous to Fig.~\ref{fig:graphene_defects}(a) for a triangular lattice. The change in the energy at the defect core is reported in Fig.~\ref{fig:stabterm}(b), shown as the energy difference with respect to the bulk $\Delta f(h,x)$, on a line crossing the defect at its core along the $[10]$ direction. Although slightly affected by the additional energy term, the shape of $\Delta f(h,x)$ remains unchanged. Furthermore, this quantity is found to depend almost linearly on $h$---as shown in the inset of Fig.~\ref{fig:stabterm}(b)---which implies that the term involving the amplitudes in Eq.~\eqref{eq:stabterm} is nearly independent of $h$. As such, only negligible changes are expected in the amplitudes when including $f_{\rm stab}$.

These observations are valid for the explored values of $h$ (up to 1 in the considered setting). Larger values may lead to more significant differences, and the smallest $h$ which results in the desired bulk phase should be used. 
Such a value may be computed by considering the bulk energy of the stripe and the targeted crystalline phase. For instance, focusing on a parameter range where crystalline phases are favored and considering a one-mode approximation with $|\mathcal{B}_j|=1$, the energy of the stripe phase in the presence of the stabilizing term reduces to $f_{\rm stripe}(\phi_{\rm s},N,h)=B\phi_{\rm s}^2+[(3/2)D+h(N-1)]\phi_{\rm s}^4$, with $\phi_{\rm s}$ the nonzero (real) amplitude minimizing the energy. Then, one can compute the bulk energy density of the targeted crystal phase $f_{\rm bulk}(\phi_{\rm c},N)$, with $\phi_{\rm c}$ the nonzero (real) amplitude minimizing the energy for the targeted crystal structure. The minimum value of $h$ is given for $f_{\rm bulk}(\phi_{\rm c},N)=f_{\rm stripe}(\phi_{\rm s},N,h)$, namely
\begin{equation}
    h_{\rm min} = \frac{f_{\rm bulk}(\phi_{\rm c},N)-B\phi_s^2-(3/2)D\phi_{\rm s}^4}{(N-1)\phi_{\rm s}^4}.
\end{equation}
The stripe phase is then penalized for $h>h_{\rm min}$. Notice that $h_{\rm min}$ is positive only for $f_{\rm stripe}(\phi_{\rm s},N,0)<f_{\rm bulk}(\phi_{\rm c},N)$, corroborating that the stabilization term is otherwise not needed. This argument can be readily extended to basis coefficients with modulus different from one and multimode approximations. For the latter case, phases characterized by having nonzero amplitudes for the first and some of the additional modes may be the lowest energy state. The stabilization term would penalize them as well. In this case, $h_{\rm min}$ should be determined by adapting the definition of $f_{\rm stripe}(\phi_{\rm s},N,h)$ to the actual lowest-energy phase.

Interestingly, the behavior illustrated in Fig.~\ref{fig:stabterm}(b) closely resembles the features of the additional energy term introduced in Ref.~\cite{Salvalaglio2017} to tune the energy of defects in APFC models. Therefore, although further explorations in this direction are beyond the scope of this work, we envisage applications of such a term in this regard.

\subsection{Elastic properties}

Elastic properties within the APFC model can be derived from the differential operator $\mathcal{M}_{\rm{M},n}$ \cite{SalvalaglioMSMSE2022} by considering deformations from the perfect lattice as encoded in the amplitudes defined as in Eq.~\eqref{eq:inieta}. Inserting this form back in Eq.~\eqref{eq:n_app_basis}, and exploiting the approximation \eqref{eq:MM}, the elastic energy density reads
\begin{equation}
\begin{split}
f_{\rm elas}
\approx 4A \sum_{n=1}^N  \Gamma_n |\mathcal{B}_n|^2\phi_n^2
 k_i^n k_j^n k_l^n k_m^n U_{ij}U_{lm},
\end{split}
\label{eq:elas_full}
\end{equation}
with $k_i^n$ the $i$th component of $\kv_n$ and $U_{ij}$ the nonlinear strain tensor \cite{Huter2016,SalvalaglioMSMSE2022}
\begin{equation}
U_{ij}=\frac{1}{2}\left(u_{ij}+u_{ji}-u_{il}u_{jl}\right),
\end{equation}
using the Einstein summation convention. We remark that in Eq.~\eqref{eq:elas_full} we neglected strain gradient terms. By writing the elastic part of the free energy as $f_{\rm elas}=(1/2)\sigma_{ij}U_{ij}$ with $\sigma_{ij}=\mathcal{C}_{ijkl}U_{kl}$ the components of the stress field and $\mathcal{C}_{ijkl}$ the elastic modulus tensor, one gets 
\begin{equation}
\mathcal{C}_{ijkl}= 8A\sum_{n=1}^N \Gamma_n |\mathcal{B}_n|^2\phi_n^2
 k_i^n k_j^n k_l^n k_m^n.
\label{eq:lambda}
\end{equation}
From this expression, the symmetry encoded in the elastic constants is dictated by the symmetry of the Bravais lattice with a prefactor
\begin{equation}
    |\mathcal{B}_n|^2=J+2\sum_i^J\sum_{j\neq i}^J \cos\left[\kv_n \cdot (\mathbf{R}_j-\mathbf{R}_i) \right],
\end{equation}
depending on the local structure, which consistently reduces to 1 for $J=1$.

\section{Numerical Examples}\label{sec::numerical}

In this section, we apply the extended APFC model to the simulation of crystalline structures beyond Bravais lattices hosting isolated defects. These simulations illustrate the possibility of describing periodic arrangements of different kinds. Moreover, they showcase the stability of the considered bulk phases since defects represent a significant perturbation leading to the nucleation and growth of more stable phases, if present. The evidence reported below also serves as proof of concept for studying edge dislocations in exotic structures and the technologically relevant diamond structure.


We first address the case of straight edge dislocations arranged in prescribed positions. We define amplitudes encoding a lattice distortion induced by dislocations exploiting Eq.~\eqref{eq:inieta} with $\mathbf{u}(\mathbf{r})=\sum_d^D \mathbf{u}^{(d)}(\mathbf{r})$ the displacement field obtained as a superposition of the displacement induced by $D$ dislocations with Burgers vector $\mathbf{b}^{(d)}\parallel \hat{\mathbf{x}}$ and dislocation line $\mathbf{l}^{(d)}\parallel \hat{\mathbf{z}}$. According to classical continuum mechanics, in the approximation of linear and isotropic elasticity, $\mathbf{u}^{(d)}(\mathbf{r})$ components read \cite{anderson2017}
\begin{equation}\label{eq:dispfield}
    \begin{split}
    u_x^{(d)}(\mathbf{r}) &= \frac{|\mathbf{b}|}{2\pi} \left[ \arctan\left(\frac{\bar{y}}{\bar{x}}\right) + \frac{\bar{x} \bar{y}}{2(1-\nu)r^2} \right],\\
    u_y^{(d)}(\mathbf{r}) &= -\frac{|\mathbf{b}|}{2\pi} \left[ \frac{(1-2\nu)\log(r^2)}{4(1-\nu)} + 
                \frac{\bar{x}^2-\bar{y}^2}{4(1-\nu)r^2} \right], \\
    u_z^{(d)}(\mathbf{r}) &= 0,
    \end{split}
\end{equation}
with $\mathbf{r}=(x,y,z)$, $r^2=\bar{x}^2 + \bar{y}^2$, $\bar{x}=x-x_0^{(d)}$, $\bar{y}=y-y_0^{(d)}$, $(x_0^{(d)},y_0^{(d)})$ the position of the $d$th straight dislocation, and $\nu=1/3$.
A squared domain with side $L$ in the $xy$ plane is considered, and we define four edge dislocations in the positions $(\pm L/4,\pm L/4)$. The modulus of the Burgers vectors is taken to be equal to one lattice spacing (depending on the specific symmetry under investigation, see below), oriented parallel and antiparallel to $\hat{\mathbf{x}}$ such that $\mathbf{b}^{(d)}\cdot \hat{\mathbf{x}}=\mathrm{sign}(x_0^{(d)}y_0^{(d)})|\mathbf{b}^{(d)}|$. Furthermore, we consider periodic boundary conditions. The interaction with periodic images is accounted for in the initial condition by considering dislocations arranged periodically in a region $11L \times11L$ entering Eq.~\eqref{eq:dispfield}. In the following, we will focus on one of the dislocations in such arrays.

\begin{figure}
    \centering
    \includegraphics{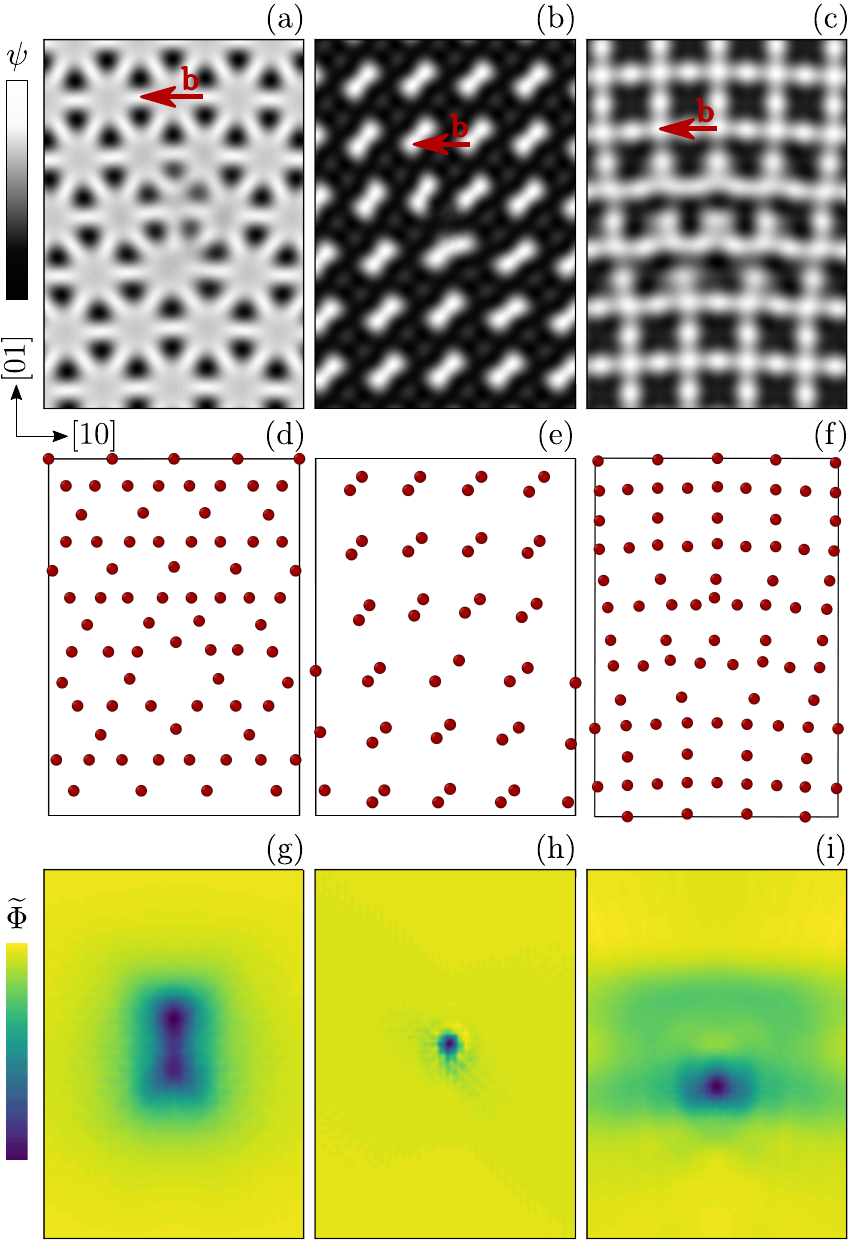}
    \caption{Edge dislocations in two-dimensional lattices with a basis. $\mathbf{k}_j$ and $\mathbf{R}_j$ are reported in the Appendices \ref{app:triangular} and \ref{app:square}.
    The density $\psi$ is plotted in the first row, and the Burgers vectors are marked in green. 
    In the second row, spheres of arbitrary radius are drawn on the density maxima. 
    In the third row, $\widetilde{\Phi} = 2\sum_{n=1}^N |\widetilde{\eta}_n|^2$ is plotted.
    [(a), (d), (g)] Kagome structure, three-mode triangular lattice with a diatomic basis; $\bar{\psi} = 0,\,A=0.02,\, B=0.02,\, C=0.5,\, D=1/3,\, b_1=0.1,\, b_2=b_3=0,\, h=0$. 
    [(b), (e), (h)] Dimer structure, three-mode square lattice with diatomic basis. $\bar{\psi} = 0,\,A=0.02,\, B=-0.1,\, C=-1,\, D=1/3,\, b_j=0,\, h=0$.
    [(c), (f), (i)] Frame structure, three-mode square lattice with triatomic basis; $\bar{\psi} = 0,\,A=0.02,\, B=-0.1,\, C=0.05,\, D=1/3,\,  b_j=0,\, h=0.2$. }
    \label{fig:edge2D}
\end{figure}

In Fig.~\ref{fig:edge2D}, we report simulation results for representative two-dimensional lattices. 
For each system, we show the field $\widetilde{\Phi}$, the reconstructed density, and, to allow for an easy visualization of the lattice, a reconstructed discrete lattice obtained by drawing spheres of arbitrary radius at the position of density maxima using the visualization tool OVITO \cite{ovito}. Details about the settings for the corresponding bulk phases can be found in Appendices~\ref{app:triangular} and \ref{app:square}, while parameters are reported in the caption.
In Figs.~\ref{fig:edge2D}(a), \ref{fig:edge2D}(d), and \ref{fig:edge2D}(g), we show an edge dislocation in the kagome structure. The magnitude of the Burgers vector is equal to the lattice parameter $a=4\pi/\sqrt{3}$, and the domain size is $L=60\,a$. A bulk phase analogous to our result was obtained by the PFC model \cite{Mkhonta2013} with settings similar to the one exploited for the results in Fig.~\ref{fig:graphene_bi}(a). The density corresponds to an inverse, three-mode graphene structure. Indeed, the parameters are those used for Fig.~\ref{fig:graphene_defects}(a), except for the opposite sign of $C$ (coefficient of the cubic term in the energy). The maxima for this phase are localized and arranged as a kagome structure, as shown in Fig.~\ref{fig:graphene_defects}(d). Importantly, the considered edge dislocation results in a $5|7$ defect core in good agreement with known results for defects in kagome structures \cite{paulose2015topological}. 
In Figs.~\ref{fig:edge2D}(b), \ref{fig:edge2D}(c), \ref{fig:edge2D}(e), \ref{fig:edge2D}(f), \ref{fig:edge2D}(h), and \ref{fig:edge2D}(i), we show edge dislocations in two square lattices with different bases. The Burgers vector is set equal to the lattice parameter $a=2\pi$, and the domain size is $L = 70\,a$. Similarly to the kagome structure discussed above, they represent exotic crystal arrangements, namely a squared arrangement of dimers aligned along the diagonal of the square unit cell [Figs.~\ref{fig:edge2D}(b) and \ref{fig:edge2D}(e)] and a triatomic basis forming a periodic arrangement of eight-atom frame structures [Figs.~\ref{fig:edge2D}(c) and \ref{fig:edge2D}(f)]. They showcase the effects of the terms introduced in the previous sections and their application to nontrivial symmetries. The structures obtained resemble exotic lattices of interest either for some materials or for periodic arrangements of other objects in general \cite{Reichhardt2002,overy2016design,kishida2022direct}.

\begin{figure}
    \centering
    \includegraphics{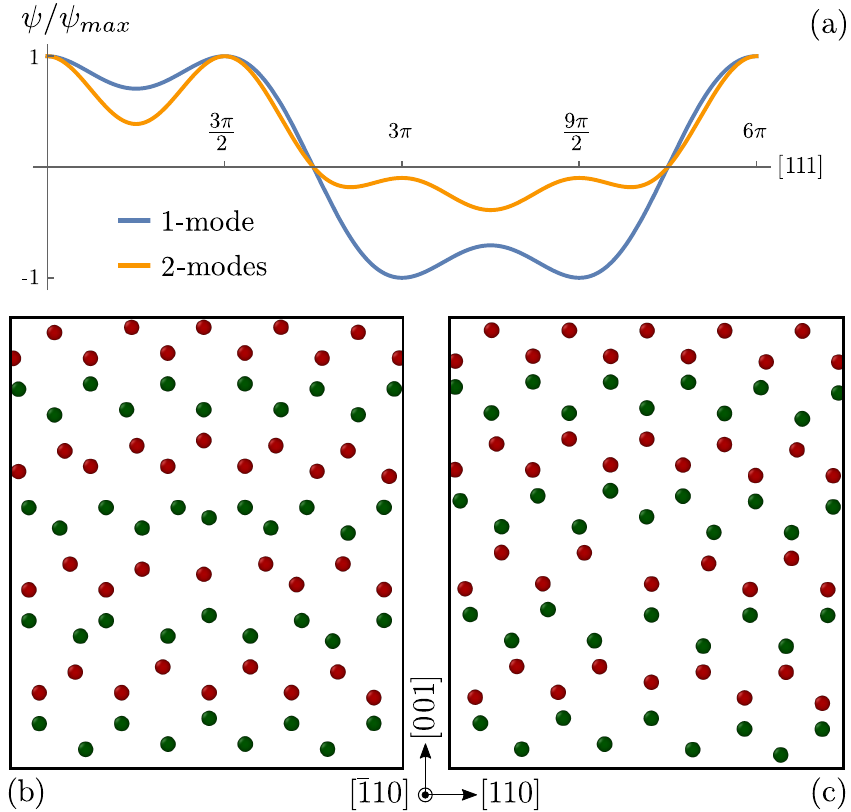}
    \caption{Diamond lattices reconstructed from APFC simulations. 
    (a) One-dimensional density profile along the diagonal of the unit cell for one- and two-mode approximations. 
    (b) Edge dislocation, one-mode approximation,  $\bar{\psi} = 0,\,A=1,\,B=-0.02,\,C=0,\,D=1/3,\,h=0.2$.  
    (c) Edge dislocation, two-mode approximation, $\bar{\psi} = 0,\,A=0.36,\,B=0.1,\,C=-1,\,D=1/3,\,h_j=0,\,b_1=0,\,b_2=0.5$. 
    Details on $\mathbf{k}_j$ and $\mathbf{R}_j$ are reported in Appendix \ref{app:diamond}. In (b) and (c), spheres of arbitrary radius are drawn at the density peaks, and two consecutive $(\bar{1}10)$ planes are shown in different colors (red and green). }
    \label{fig:dia_lomer}
\end{figure}

The amplitude expansion enables the study of relatively large, three-dimensional systems \cite{Praetorius2019,SalvalaglioNPJ2019}. We focus now on the novel APFC modeling of a three-dimensional lattice of great relevance for physical applications: the diamond lattice. 
It can be defined by introducing a diatomic basis in the fcc lattice \cite{ashcroft1976solid}. Details on the specific choice of $\mathbf{k}_j$ and $\mathbf{R}_j$ are reported in Appendix \ref{app:diamond}. 
In the one-mode approximation, the diamond structure can be obtained as the most stable phase by exploiting the stabilization term introduced in Sec.~\ref{sec:3c}, as stripe phases are more stable otherwise within the whole parameter range. It is worth mentioning that this does not include $C$ as no cubic terms result from Eq.~\eqref{eq:pol_resonance}. Conversely, a degree-three term is present in the two-mode approximation, and relatively large $|C|$ values allow for obtaining the diamond-lattice phase as the most stable one, even without the stabilization term. This reflects the stability of the underlying Bravais (fcc) lattice \cite{ElderPRE2010,Salvalaglio2017}.
A one-dimensional density plot along the $[111]$ direction is shown in Fig.~\ref{fig:dia_lomer}(a) for the one-mode and two-mode approximations, respectively. Both show two distinct peaks at 0 and $\frac{3}{2}\pi$, corresponding to the two atoms in the basis. In the one-mode approximation, the minimum at $\frac{3}{4}\pi$ is relatively shallow. This can be improved by considering a two-mode approximation as shown in Fig.~\ref{fig:dia_lomer}(b). In the two-mode approximation, additional small features develop, such as two relative maxima at $3\pi$ and $\frac{9}{2}\pi$, which can be controlled by varying the multimode coefficients $b_i$.

In Figs.~\ref{fig:dia_lomer}(b) and \ref{fig:dia_lomer}(c), we show edge dislocations in a diamond lattice. We consider pure-edge dislocations, with Burgers vector $\mathbf{b} = \pm \frac{a}{2}[110](001)$, where $a=2\sqrt3\pi$ is the side of the cubic unit cell. We take a slab-shaped domain, with dimensions $L_x=L_y=40\,a,\,L_z=4\,a$. 
For the sake of clarity, we draw spheres corresponding to the density peaks to visualize the reconstructed density (now a function of three spatial coordinates) and only display two $(\bar{1}10)$ planes, marked in red and green.
In Figs.~\ref{fig:dia_lomer}(b) and \ref{fig:dia_lomer}(c), we show a relatively small region centered on one dislocation obtained by one-mode and two-mode approximations, respectively. In both cases, the diamond lattice represents a stable phase while hosting the significant lattice distortion induced by dislocations. Moreover, even though we solve for slowly varying variables (the amplitudes), a dislocation core corresponding to a ring of eight atoms is obtained in the reconstructed density, which is consistent with results derived in literature by geometrical arguments \cite{Hornstra1958}. The two-mode approximation in Fig.~\ref{fig:dia_lomer}(c) delivers a better approximation of such known dislocation-core structure, thus confirming the higher resolution achieved with multimode approximations.

\begin{figure}
    \centering
    \includegraphics{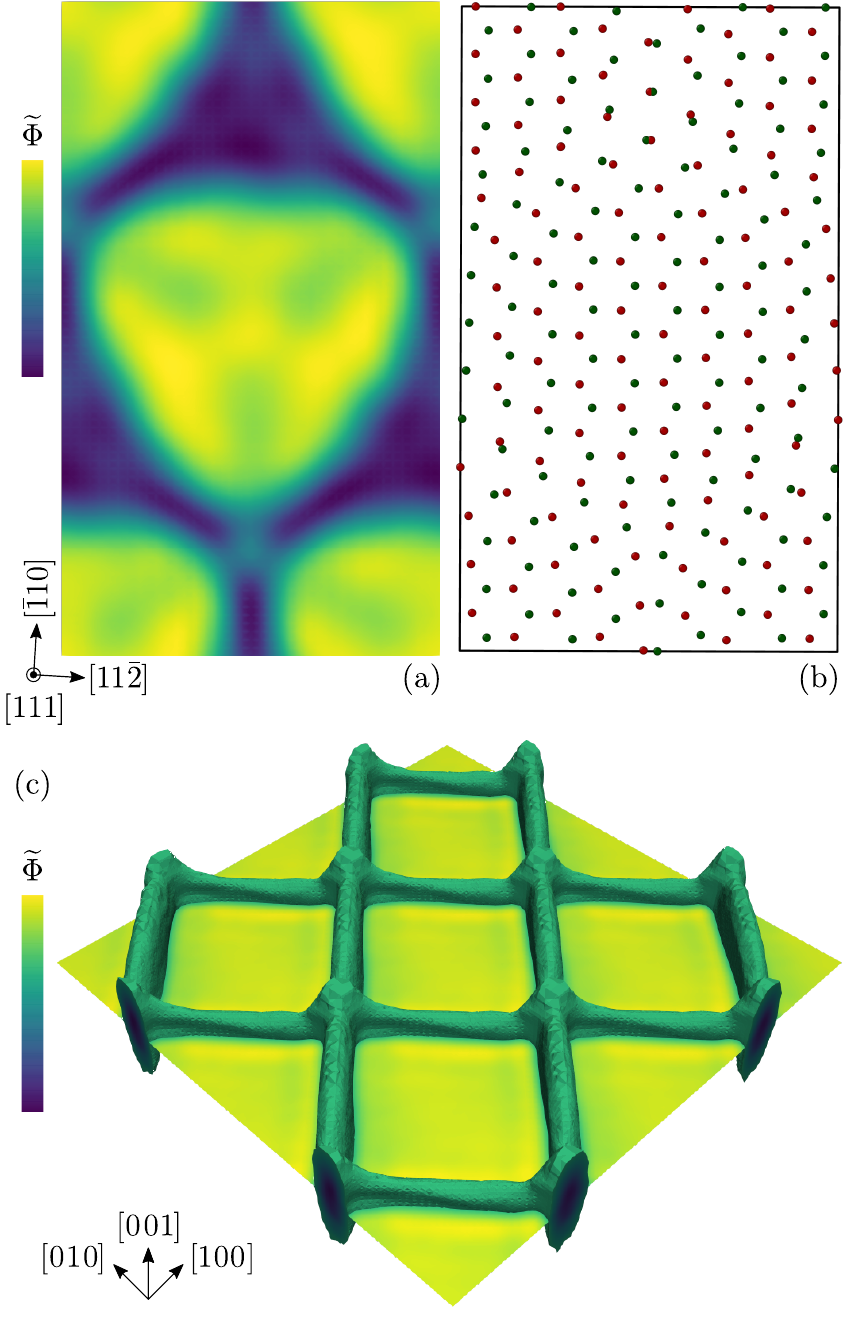}
    \caption{
    Small-angle twist grain boundaries in the diamond lattice. A one-mode approximation is considered with $\bar{\psi} = 0,\,A=1,\,B=-0.02,\,C=0,\,D=1/3,\,h=0.2$. 
    [(a) and (b)] Rotation about $\mathbf{\hat{z}} = [111]$ by $\theta=7.34^\circ$, corresponding to the $\Sigma183$ coincidence site lattice. The coordinate system is rotated by $\theta/2$.
    Panel (a) shows $\widetilde{\Phi} = 2\sum_{n=1}^N |\widetilde{\eta}_n|^2$ at the interface. 
    Panel (b) shows spheres of arbitrary radius placed at the density peaks. Two consecutive (111) planes are shown in different colors.
    (c) Rotation about $\mathbf{\hat{z}} = [001]$ by $\theta=7.34^\circ$. Plot of $\widetilde{\Phi}$ at the interface showing the dislocation network in three dimensions. This is obtained as the region for which $\widetilde{\Phi}<0.7\,\widetilde{\Phi}_{\rm max}$.
    }
    \label{fig:dia_twistGB}
\end{figure}

As a final example, we consider three-dimensional small-angle grain boundaries in the diamond structure (Fig.~\ref{fig:dia_twistGB}) as prototypical cases in which dislocation form due to mismatch or misorientation of crystalline structures meeting at an interface, rather than being introduced explicitly as for the cases above. We consider the diamond lattice representation achieved with the one-mode approximation, and we define a small-angle rotation about the [111] direction, set as the $z$-axis of the simulation domain. 
We define a rotated lattice in the $z>0$ half-space,
while leaving the $z<0$ half-space unrotated. Periodic boundary conditions are applied so that two twist grain boundaries form, namely at the interface between the two portions of the crystal and at the boundaries with normal along $\hat{\mathbf{z}}$.
In the rotated lattice, the amplitudes can be expressed as
\begin{equation}
    \eta_n = \phi_n e^{{\rm i}\delta \mathbf{k}_n\cdot\mathbf{r}},
\end{equation}
where $\phi_n$ is the value of the amplitude in the relaxed bulk, and $\delta \mathbf{k}_n=\mathcal{R}(\theta)\mathbf{k}_n-\mathbf{k}_n$ is the difference between the reciprocal-lattice vectors in the rotated crystal with respect to the reference one \cite{Salvalaglio2018}, with $\mathcal{R}(\theta)$ the counterclockwise rotation matrix about the $z$ direction.
In the rotated crystal, the amplitudes oscillate in the $x,\,y$ plane with wavelengths $[\lambda_n]_{x,y} = 2\pi/[\delta k_n]_{x,y}$. 
The unit cell of the periodic dislocation pattern forming at the interface is obtained by rotating the whole domain about $\hat{\mathbf{z}}$ such that the shortest vector $\delta \mathbf{k}_n$ is aligned/perpendicular with two boundaries \cite{Salvalaglio2018}, namely corresponding to a rotation of $-\theta/2$.

When considering rotations about the [111] axis, we take an angle $\theta = 7.34^\circ$, which is expected to result in a $\Sigma183$ coincidence site lattice \cite{Baruffi2021}. The domain size is $L_x \approx 6\,a ,\,L_y \approx 10\,a$, where $a=2\sqrt{3}\pi$ is the side of the cubic unit cell, and we choose arbitrarily $L_z = 2L_y$. 
In Fig.~\ref{fig:dia_twistGB}(a), we show $\widetilde{\Phi}$, which decreases at the dislocations, as discussed in Sec. \ref{sec:3a}. Here we observe a pattern with localized dislocation cores extending a few ($\sim$3) lattice spacings. A hexagonal dislocation network enclosing regions of local coherency forms.
To further appreciate the separation between coherent and noncoherent regions, in Fig.~\ref{fig:dia_twistGB}(b) we draw spheres of arbitrary radius at the density peaks for two consecutive (111) planes, represented by different colors. In the central part of the hexagonal region bounded by the dislocation network, a coherent region featuring diamond crystal structure is present. The coherency is lost at the dislocation network.   
This structure agrees very well with recent results obtained by molecular dynamics simulations for perfect shuffle dislocations in Si \cite{Baruffi2021}.
Importantly, the small-angle grain boundary reported is peculiar to the diamond structure and deviates from the ones obtained with an analogous setting for an fcc lattice \cite{Salvalaglio2018} (i.e., with one atom per Bravais lattice site). This further assesses the description of lattices with basis (Sec.~\ref{sec:3b}) we introduced. 

Another representative dislocation network is obtained by considering a rotation about the $[001]$ axis, here simulated with the same angle $\theta=7.34^\circ$ considered before for the sake of simplicity. The domain size for this case reads $L_x=L_y\approx8\,a,\,L_z=2L_x$. The result is reported in Fig.~\ref{fig:dia_twistGB}(c), which shows the formation of a dislocation network featuring dislocation lines along $[110]$ and $[\bar{1}10]$, thus corresponding to the ones illustrated in Fig.~\ref{fig:dia_lomer}. To appreciate the periodicity of the dislocation network, the simulation cell is repeated $2\times2$ times. It resembles dislocation patterns obtained in heterostructures with diamond lattice symmetries and $(001)$ interfaces featuring a mismatch and/or twist among layers as observed, e.g., in conventional group-IV epitaxy and heteroepitaxy \cite{Paul_2004,AQUA201359}.

\begin{figure}
    \centering
    \includegraphics{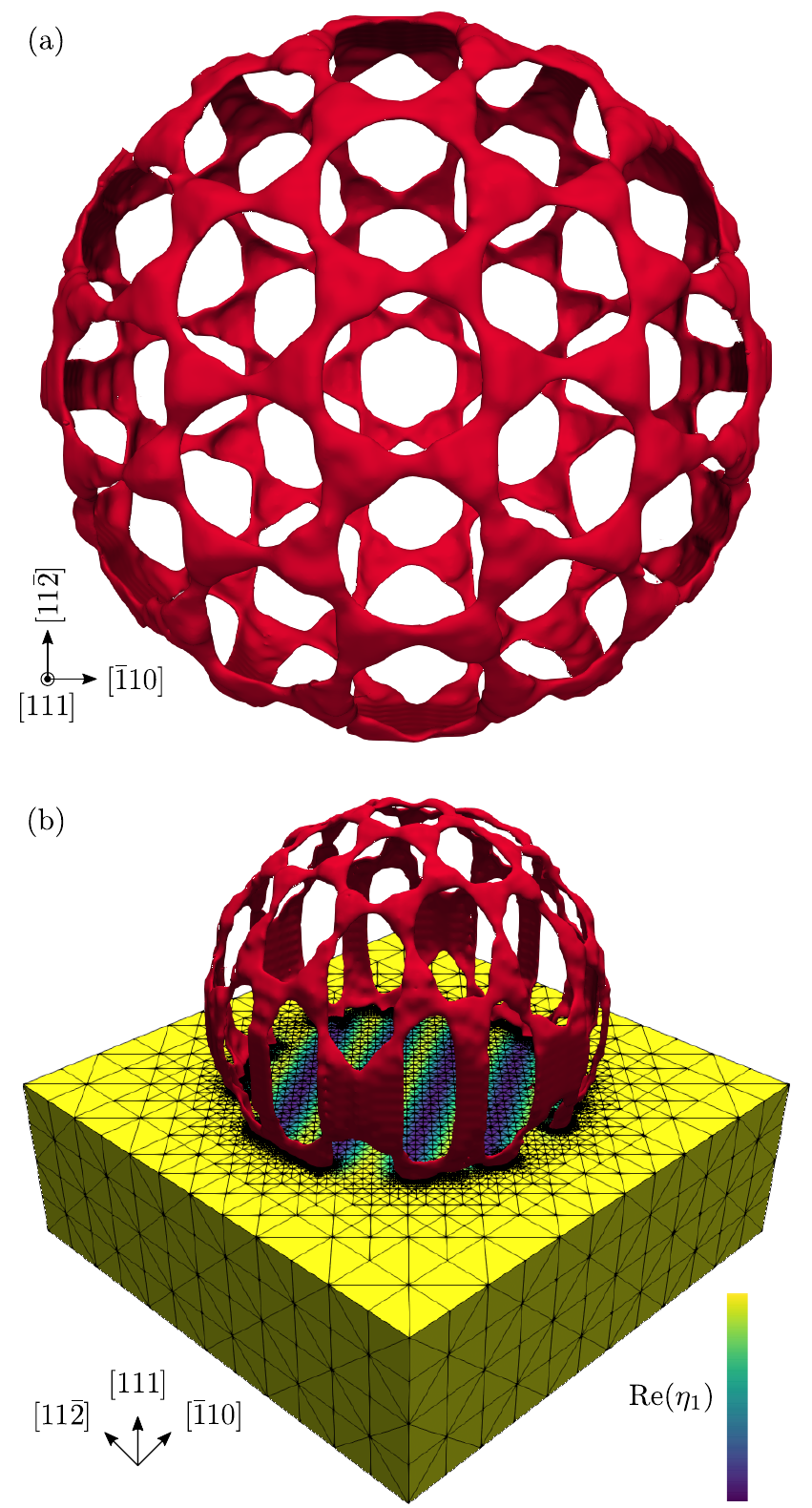}
    \caption{Spherical inclusion in the diamond lattice, rotated about the [111] axis by an angle of $3^\circ$. The sphere radius is 34 lattice spacings. 
    A two-mode approximation is considered with $\bar{\psi} = 0,\,A=0.36,\,B=0.1,\,C=-1,\,D=1/3,\,h_j=0,\,b_1=0,\,b_2=0.5$.
    (a) Defect structure, obtained as the region of space in which $\widetilde{\Phi}<0.7\,\widetilde{\Phi}_{\rm max}$. Arbitrary uniform color is used.
    (b) Defect structure and cutaway of the simulation box. Re$(\eta_1)$ is plotted on the box surface. The locally refined simulation grid is shown. }
    \label{fig:dia_inclusion}
\end{figure}
    
Finally, to showcase larger three-dimensional systems that can be handled with the proposed approach, we considered a spherical inclusion rotated about the $[111]$ direction in a diamond lattice, modeled here through a two-mode approximation as in Fig.~\ref{fig:dia_lomer}(c). The radius of the inclusion is set to 34 lattice spacings, embedded in a cubic matrix with a size of 96 lattice spacings. In terms of physical units, Si has a relaxed lattice parameter $a\approx 0.54$~nm at 300~K \cite{codata2018}, meaning that the side of the simulation box has a length $L\approx 51.84$~nm, and the sphere a radius $r\approx18.36$~nm. 
Consistently with the previous example shown in Fig.~\ref{fig:dia_twistGB}, the interface between the inclusion and the matrix with normal along the [111] shows a defect structure featuring extended defect cores with a hexagonal arrangement of triangularlike regions with loss of coherency, as illustrated in Fig.~\ref{fig:dia_inclusion}(a). This structure connects with tiltlike (small-angle) grain boundaries when the normal to the interface deviates from the $[111]$ direction, complementing previously reported evidence for basic lattice symmetries \cite{Yamanaka2017,Salvalaglio2018}. 
In Fig.~\ref{fig:dia_inclusion}(b) we report an illustration of the mesh used, superimposed to a plot of Re$(\eta_1)$. We remark that the mesh is coarse in the matrix, where the amplitudes are constant, refined inside the inclusion, where the amplitudes oscillate slowly, and even more refined around the defect structure, where the amplitudes vary quickly. This showcases in a practical, large-scale case the benefit of considering real-space methods in terms of inhomogeneous and optimized spatial discretization.

\section{Conclusions}\label{sec:conclusions}

We extended the coarse-grained description of crystal lattices delivered by the APFC model. A formulation suitable for multimode approximations of crystalline atomic densities with an adjustable resolution and an extended parametrization was introduced, enabling the description of a wide range of lattice structures. It was shown to recover the main features of the corresponding microscopic PFC model for two-dimensional lattices \cite{Mkhonta2013}. 
Moreover, the amplitude expansion of Bravais lattices with a basis was derived, significantly extending the capability of the model toward the study of systems of general interest. A stabilization term was introduced to favor crystal structures when competing with stripelike phases \cite{Elder2002}. Notice that the proposed framework goes beyond the specific model formulation considered here. It may be exploited with other formulations for the energy functional, such as the amplitude expansion of the XPFC model \cite{Ofori-Opoku2013}. We also envisage applications of the concepts introduced in this work to coarse-grained formulations of alternative microscopic models in which the order parameter may be expanded in Fourier modes, e.g., theories based on $\ln(\psi)$ rather than $\psi$ recently proven to provide improved accuracy of microscopic lengthscales and related elastic properties \cite{Archer2019,Subramanian2021,Ganguly2022}.

Examples were shown for the renowned kagome lattice, square-lattice arrangements of dimers and trimers, and the diamond lattice, all hosting dislocations. Excellent agreement is obtained with dislocation networks known for such lattices, approaching the description obtained with atomistic models.
These examples thus demonstrate the stability of the phases described by the extended APFC model while delivering proofs of concept for describing dislocations and lattice deformations therein.
Furthermore, the description of the diamond structure is expected to enable the mesoscale treatment of technology-relevant systems such as group-IV semiconductors (Si and Ge), which are in high demand for the study of nanostructures for optoelectronics applications \cite{Paul_2004,AQUA201359}. 

The proposed formulation is suitable for real-space numerical methods. Therefore, it allows for exploiting adaptive meshes \cite{Athreya2006,Praetorius2019,Bercic2018}, which have
been one of the first motivations for devising the APFC model from PFC, and are essential to reach large scales \cite{SalvalaglioMSMSE2022}.
In this context, it is also relevant that the equations from the proposed extensions, particularly the multimode APFC model, feature the same differential order (fourth) as in the classical one-mode APFC model, thus introducing no additional stiffness or complexity. 
Still, we remark that the number of equations to solve depends on the number of reciprocal-space vectors in the amplitude expansion. Therefore, the number of modes to be considered must be a trade-off between the desired resolution and the computational cost. Even so, we note that in the context of microscopic theories, complex amplitudes are usually adopted for (semi-)analytical derivations of, e.g., dislocation velocities and the stress field \cite{SalvalaglioMSMSE2022}, and need to be derived from the density field. By contrast, the proposed APFC model may be readily used to analyze such theoretical aspects within the broad context of PFC modeling, as it solves directly for the complex amplitudes and can reproduce complex lattice symmetries.

We focused here on the description of the lattice symmetry by the APFC model through extensions, while leaving the structure of the equations and the meaning of the quantities defined therein unchanged. This would then allow combining such descriptions with recent developments \cite{SalvalaglioMSMSE2022} focusing, e.g., on the advanced description of elasticity \cite{HeinonenPRL2016,SalvalaglioJMPS2020} as well as binary \cite{ElderPRE2010,SalvalaglioPRL2021} and multiphase systems \cite{Ofori-Opoku2013}. Another interesting perspective is the combination of the versatility achieved here for crystalline structures in the solid phase with a more general description of phases, including liquid and vapor phases (see, e.g., Ref.~\cite{Jreidini2022}).

\section*{Acknowledgements}
The authors acknowledge useful discussions with Ken R. Elder and Axel Voigt.
This work has been funded by the Deutsche Forschungsgemeinschaft (DFG – German Research Foundation), Grant No.~SA4032/2-1. Computing resources have been provided by the Center for Information Services and High-Performance Computing (ZIH) at TU Dresden.

\appendix

\section{Triangular lattices}\label{app:triangular}
The reciprocal space vectors used for the one-mode approximation are
\begin{equation}\label{eq:kvec_tri_1mode}
\begin{split}
        \mathbf{k}_1 &= \left(0,1\right),\,
        \\ \mathbf{k}_2 &= \left(-\frac{\sqrt{3}}{2},-\frac{1}{2}\right),\,
        \\ \mathbf{k}_3 &= \left(\frac{\sqrt{3}}{2},-\frac{1}{2}\right).
        \end{split}
\end{equation}
In addition, for the three-mode approximation, the following reciprocal-space vectors are considered:
\begin{equation}
\begin{split}
    \mathbf{k}_4 &= \mathbf{k}_3 - \mathbf{k}_2,\\
    \mathbf{k}_7 &= 2\mathbf{k}_1,
\end{split}
\quad
\begin{split}
    \mathbf{k}_5 &= \mathbf{k}_1 - \mathbf{k}_3,\\
    \mathbf{k}_8 &= 2\mathbf{k}_2,
\end{split}
\quad
\begin{split}
    \mathbf{k}_6 &= \mathbf{k}_2 - \mathbf{k}_1,\\
    \mathbf{k}_9 &= 2\mathbf{k}_3.
\end{split}    
\end{equation}
The triangular lattice does not require a basis, i.e. $\mathcal{B}_n=1$. For the honeycomb and kagome lattices, two atoms are considered in the unit cell, with positions
\begin{equation}
    R_1 = \left(0,0\right),\quad
    R_2 = \frac{4}{3}\pi\left(0,1\right),
\end{equation}
resulting in the following values for the $\mathcal{B}_n$ coefficients:
\begin{equation}
\begin{split}
    \mathcal{B}_{1,2,3} &= \frac{1}{2}+i\frac{\sqrt{3}}{2},\\
    \mathcal{B}_{4,5,6} &= 2,\\
    \mathcal{B}_{7,8,9} &= \frac{1}{2}-i\frac{\sqrt{3}}{2}.
\end{split}
\end{equation}

\section{Square lattices}\label{app:square}
The square lattice is obtained by a two-mode approximation with reciprocal lattice vectors
\begin{equation}
\begin{split}
    \mathbf{k}_1 &= \left(1,0\right),\\
    \mathbf{k}_3 &= \left(1,1\right),
\end{split}
\quad
\begin{split}
    \mathbf{k}_2 &= \left(0,1\right),\\
    \mathbf{k}_4 &= \left(1,-1\right).
\end{split}
\end{equation}

For the dimer structure as in Fig.~\ref{fig:edge2D}(b,e,h), we consider an additional mode with reciprocal lattice vectors
\begin{equation}
\begin{split}
    \mathbf{k}_5 &= \left(1,2\right),\\
    \mathbf{k}_7 &= \left(2,-1\right),
\end{split}
\quad
\begin{split}
    \mathbf{k}_6 &= \left(2,1\right),\\
    \mathbf{k}_8 &= \left(-1,2\right),
\end{split}
\end{equation}
and a diatomic basis aligned with the diagonal of the unit cell:
\begin{equation}
    R_1 = \left(0,0\right),\quad
    R_2 = \left(\frac{\pi}{2},\frac{\pi}{2}\right),
\end{equation}
This choice results in the following basis coefficients:
\begin{equation}
    \mathcal{B}_{1,2,7,8} = 1-i,\quad
    \mathcal{B}_{3} = 0,\quad
    \mathcal{B}_{4} = 2,\quad
    \mathcal{B}_{5,6} = 1+i.
\end{equation}
Notably, since $\mathcal{B}_{3} = 0$, $\mathbf{k}_3$ does not contribute to the description of the lattice, and can be omitted entirely.

For the frame structure as in Fig.~\ref{fig:edge2D}(c,f,i), we consider a different third mode, with reciprocal lattice vectors
\begin{equation}
    \mathbf{k}_5 = 2\mathbf{k}_1, \quad
    \mathbf{k}_6 = 2\mathbf{k}_2,
\end{equation}
and a triatomic basis, with one atom in the corner and two atoms in the center of the side of the unit cell 
\begin{equation}
    R_1 = \left(0,0\right),\quad
    R_2 = \left(\pi,0\right),\quad
    R_3 = \left(0,\pi\right),
\end{equation}
resulting in the following basis coefficients:
\begin{equation}
    \mathcal{B}_{1,2} = 1,\quad
    \mathcal{B}_{3,4} = -1,\quad
    \mathcal{B}_{5,6} = 3.
\end{equation}

\section{Diamond lattices}\label{app:diamond}
The diamond lattice consists of a fcc crystal structure with a diatomic basis. 
Thanks to the stabilizing term from Eq.~\eqref{eq:stabterm}, it is possible to describe the diamond lattice with a one-mode approximation, using the following reciprocal-space vectors:
\begin{equation}
\begin{split}
    \mathbf{k}_1 &= k_0\left(-1,1,1\right), \\
    \mathbf{k}_3 &= k_0\left(1,1,-1\right),
\end{split}    
\qquad
\begin{split}
    \mathbf{k}_2 &= k_0\left(1,-1,1\right), \\
    \mathbf{k}_4 &= k_0\left(-1,-1,-1\right),
\end{split}
\end{equation}
with $k_0 = 1/\sqrt3$. The positions of the two atoms within the unit cells are taken to be
\begin{equation}
    R_1 = \left(0,0,0\right),\quad
    R_2 = \frac{\sqrt3}{2}\pi\left(1,1,1\right),
\end{equation}
resulting in the basis coefficients having the same complex value $\mathcal{B}_{1,2,3,4} = 1-i$.

The second shortest length in the reciprocal lattice is accounted for by the following reciprocal-space vectors:
\begin{equation}
\begin{split}
    \mathbf{k}_{5} &= \mathbf{k}_1 + \mathbf{k}_4,
    \\ \mathbf{k}_{6} &= \mathbf{k}_2 + \mathbf{k}_4,
    \\ \mathbf{k}_{7} &= \mathbf{k}_3 + \mathbf{k}_4.
\end{split}
\end{equation}
However, with this choice the basis coefficients are $\mathcal{B}_{5,6,7} = 0$, meaning that the second mode does not contribute to the description of the diamond lattice. The next mode that does contribute is described by the reciprocal lattice vectors
\begin{equation}
\begin{split}
    \mathbf{k}_{8,9} &= \mathbf{k}_5 \pm \mathbf{k}_6,
    \\ \mathbf{k}_{10,11} &= \mathbf{k}_6 \pm \mathbf{k}_7,
    \\ \mathbf{k}_{12,13} &= \mathbf{k}_7 \pm \mathbf{k}_5.
    \end{split}
\end{equation}
The basis coefficients for the third mode are all equal: $\mathcal{B}_{8,9,10,11,12,13} = 2$. 
As reported in the main text, the diamond lattice may be described without the stabilizing term using the first and the third mode.

\end{document}